\documentclass[12pt,iop]{emulateapj}
\usepackage{natbib}

\newcommand{\rmn}[1] {{\rm #1}}
\newcommand{\sz}{Sunyaev--Zel'dovich}

\newcommand{\sza}{Sunyaev--Zel'dovich Array}

\newcommand{\eg}{e.g.,}

\newcommand{\spitzer}{{\it Spitzer}}
\newcommand{\chandra}{{\it Chandra}}
\newcommand{\hst}{{\it HST}}

\newcommand{\zphot}{z_{\mbox{\scriptsize{phot}}}}

\newcommand{\msun}{$M_\odot$}

\newcommand{\cl}{IDCS~J1426.5+3508}

\newcommand{\rfive}{r_{\mbox{\scriptsize 500}}}

\newcommand{\Mfive}{M_{\mbox{\scriptsize 500}}}
\newcommand{\Mfivec}{M_{\mbox{\scriptsize 500,c}}}

\newcommand{\Mtwoc}{M_{\mbox{\scriptsize 200,c}}}
\newcommand{\Mtwom}{M_{\mbox{\scriptsize 200,m}}}

\def\spose#1{\hbox to 0pt{#1\hss}}
\def\simlt{\mathrel{\spose{\lower 3pt\hbox{$\mathchar"218$}}
     \raise 2.0pt\hbox{$\mathchar"13C$}}}
\def\simgt{\mathrel{\spose{\lower 3pt\hbox{$\mathchar"218$}}
     \raise 2.0pt\hbox{$\mathchar"13E$}}}

\slugcomment{ApJ, in press}

\shorttitle{\cl: SZ Measurement of a Massive IR-selected Cluster at
$z=1.75$} \shortauthors{Brodwin et al.}

\newcommand{\UMKC}{1}
\newcommand{\CfA}{2}
\newcommand{\UFlorida}{3}
\newcommand{\Davis}{4}
\newcommand{\LLNL}{5}
\newcommand{\Kavli}{6}
\newcommand{\UChicago}{7}
\newcommand{\Steward}{8}
\newcommand{\UChicagoPhysics}{9}
\newcommand{\NOAO}{10}
\newcommand{\JPL}{11}
\newcommand{\Marshall}{12}

\begin{document}


\title{\cl: Sunyaev--Zel'dovich Measurement of a Massive IR-selected
  Cluster at $z=1.75$}


\author{M.~Brodwin,\altaffilmark{\UMKC,\CfA}
 A.~H.~Gonzalez,\altaffilmark{\UFlorida}
 S.~A.~Stanford,\altaffilmark{\Davis,\LLNL} 
 T.~Plagge,\altaffilmark{\Kavli,\UChicago}
 D.~P.~Marrone,\altaffilmark{\Steward}
 J.~E.~Carlstrom\altaffilmark{\Kavli,\UChicago,\UChicagoPhysics}
 A.~Dey,\altaffilmark{\NOAO}
 P.~R.~Eisenhardt,\altaffilmark{\JPL}
 C.~Fedeli,\altaffilmark{\UFlorida} 
 D.~Gettings,\altaffilmark{\UFlorida} 
 B.~T.~Jannuzi,\altaffilmark{\NOAO}
 M.~Joy,\altaffilmark{\Marshall}
 E.~M.~Leitch,\altaffilmark{\Kavli,\UChicago}
 C.~Mancone,\altaffilmark{\UFlorida}
 G.~F.~Snyder,\altaffilmark{\CfA}
 D.~Stern,\altaffilmark{\JPL} 
and G.~Zeimann\altaffilmark{\Davis}
 }


\altaffiltext{\UMKC}{Department of Physics and Astronomy, University of Missouri, 5110 Rockhill Road, Kansas City, MO 64110}
\altaffiltext{\CfA}{Harvard-Smithsonian Center for Astrophysics, 60 Garden Street, Cambridge, MA}
\altaffiltext{\UFlorida}{Department of Astronomy, University of Florida, Gainesville, FL 32611}
\altaffiltext{\Davis}{Department of Physics, University of California, One Shields Avenue, Davis, CA 95616}
\altaffiltext{\LLNL}{Institute of Geophysics and Planetary Physics, Lawrence Livermore National Laboratory, Livermore, CA 94550}
\altaffiltext{\Kavli}{Kavli Institute for Cosmological Physics, University of Chicago, Chicago, IL 60637}
\altaffiltext{\UChicago}{Department of Astronomy and Astrophysics, University of Chicago, Chicago, IL 60637}
\altaffiltext{\Steward}{Steward Observatory, University of Arizona, 933 North Cherry Avenue, Tucson, AZ 85721}
\altaffiltext{\UChicagoPhysics}{Dept.~of Physics/Enrico Fermi
  Institute, University of Chicago, Chicago, IL 60637}
\altaffiltext{\NOAO}{NOAO, 950 North Cherry Avenue, Tucson, AZ 85719}
\altaffiltext{\JPL}{Jet Propulsion Laboratory, California Institute of Technology, Pasadena, CA 91109}
\altaffiltext{\Marshall}{Department of Space Science, VP62, NASA Marshall Space Flight Center, Huntsville, AL 35812, USA}

\begin{abstract}

  We report 31 GHz CARMA observations of \cl, an infrared-selected
  galaxy cluster at $z = 1.75$.  A \sz\ decrement is detected towards
  this cluster, indicating a total mass of $\Mtwom = (4.3 \pm 1.1)
  \times 10^{14}$ \msun\ in agreement with the approximate X-ray mass
  of $\sim 5 \times 10^{14}$\msun.  \cl\ is by far the most distant
  cluster yet detected via the \sz\ effect, and the most massive $z
  \ge 1.4$ galaxy cluster found to date.  Despite the mere $\sim 1\%$
  probability of finding it in the 8.82 deg$^2$ IRAC Distant Cluster
  Survey, \cl\ is not completely unexpected in $\Lambda$CDM once the
  area of large, existing surveys is considered.  \cl\ is, however,
  among the rarest, most extreme clusters ever discovered, and indeed
  is an evolutionary precursor to the most massive known clusters at
  all redshifts.  We discuss how imminent, highly sensitive \sz\
  experiments will complement infrared techniques for statistical
  studies of the formation of the most massive galaxy clusters in the
  $z>1.5$ Universe, including potential precursors to \cl.

\end{abstract}



\keywords{galaxies: clusters: individual (\cl) --- galaxies: clusters:
  intracluster medium --- galaxies: evolution --- cosmology:
  observations --- cosmology: cosmic background radiation}


\section{Introduction}
\label{Sec: introduction}

As the most massive collapsed objects, galaxy clusters provide a
sensitive probe of the cosmological parameters that govern the
dynamics of the Universe.  Traditional cluster cosmology experiments
constrain these parameters by comparing predicted cluster mass
functions with observed cluster counts \citep[for a review,
see][]{allen11}.  

The recent discovery of a handful of very massive, high redshift
($z>1$) galaxy clusters \citep{rosati09,brodwin10,foley11} has
prompted several studies of the power of individual massive, rare
clusters to probe the physics of inflation, in particular
non-Gaussianity in the density fluctuations that seed structure
formation \citep{dalal08,brodwin10,holz10,mortonson11,cayon11,
  hoyle11,enqvist11,foley11,williamson11,hotchkiss11,paranjape11}.
These rare, massive clusters were all identified from their hot
intracluster medium (ICM), whether via X-ray emission or from
prominent \sz\ (SZ) decrements.

Conversely, the most successful method for discovering large samples
of high redshift galaxy clusters, out to at least $z=1.5$, is via
infrared selection \citep[\eg][]{eisenhardt08,muzzin08}.  For
instance, the \spitzer/IRAC Shallow Cluster Survey
\citep[ISCS,][]{eisenhardt08}, which employs a full three-dimensional
wavelet search technique on a 4.5$\mu$m-selected, stellar mass-limited
galaxy sample, has discovered 116 candidate galaxy clusters and groups
at $z>1$. More than 20 of these have now been spectroscopically
confirmed at $1<z<1.5$
(\citealt{stanford05,elston06,brodwin06_iss,eisenhardt08,brodwin11}).

The cosmological importance of individual galaxy clusters depends
primarily on their rarity, with the most massive clusters at any
redshift providing the most leverage.  The maximum predicted cluster
mass increases with time due to the growth of structure.
Consequently, clusters with moderate masses, by present-day standards,
of a few $\times 10^{14}$ \msun\ are cosmologically interesting at the
highest redshifts (\eg\ $z>1.5$).  Only one such cluster, XDCP
J0044.0-2033 at $z = 1.58$ \citep{santos11}, has been found to date,
although at such a high redshift its $L_X$--based mass is quite
uncertain.  The most sensitive, large angle, ICM-based cluster survey
to date, the 2500 deg$^2$ South Pole Telescope Survey
\citep[SPT;][]{carlstrom11}, would only be able to detect such high
redshift clusters if they were more massive than $ \approx 3 \times
10^{14}$ \msun\ \citep{vanderlinde10,andersson11}.  A population of
such high-redshift clusters, if massive enough, could provide a
powerful opportunity to study the physics of the early Universe and,
in particular, to either confirm or falsify the predictions of the
$\Lambda$CDM paradigm. They would also be invaluable probes of the
early formation of massive galaxies in the richest environments.
 
To extend the ISCS infrared-selected cluster sample to the $z>1.5$
regime, we carried out the \spitzer\ Deep, Wide--Field Survey
\citep[SDWFS,][]{ashby09}, which quadrupled the \spitzer/IRAC exposure
time, leading to a survey more than a factor of 2 deeper than the
original IRAC Shallow Survey \citep{eisenhardt04}.  A new cluster
search, the \spitzer/IRAC Distant Cluster Survey (IDCS), to be
described in detail in forthcoming papers, was carried out using new
photometric redshifts calculated for the larger, deeper SDWFS sample.
Cluster \cl\ at $z=1.75$ \citep[][hereafter S12]{stanford12} is the
first of the new high--redshift IDCS clusters to be spectroscopically
confirmed.

In this paper we report the robust detection of the SZ decrement
associated with \cl\ in 31 GHz interferometric data taken with the
\sza\ (SZA), a subarray of the Combined Array for Research in
Millimeter-wave Astronomy (CARMA).  This is by far the most distant
cluster yet detected in the SZ effect, a consequence of the fact that
\cl\ is the most massive cluster yet discovered at $z>1.4$.  In
\textsection{\ref{Sec:Cluster}} we briefly describe the discovery of
cluster \cl.  In \textsection{\ref{Sec:Data}} we present the CARMA
observations and reductions, and describe the measurement of the
Comptonization and mass of \cl.  We also discuss various systematic
uncertainties present in our analysis.  In
\textsection{\ref{Sec:Rarity}}, we calculate the likelihood of having
found this cluster, given our survey selection function, in a standard
$\Lambda$CDM cosmology.  In \textsection{\ref{Sec:Evolution}}, we
place \cl\ in an evolutionary context, finding that it is a precursor
to the most massive known clusters at all redshifts.  In
\textsection{\ref{Sec:Discussion}} we discuss our results in light of
current and upcoming SZ surveys.  We present our conclusions in
\textsection{\ref{Sec:Conclusions}}.  In this paper, we adopt the
WMAP7 cosmology of $(\Omega_\Lambda, \Omega_M, h) = (0.728, 0.272,
0.704)$ of \citet{komatsu11}.  Except where otherwise specified (in
\textsection{\ref{Sec:Cluster}} and
\textsection{\ref{sec:sza_analysis}}), we use a cluster mass
$(\Mtwom)$ defined as the mass enclosed within a spherical region of
mean overdensity $200 \times \rho_{\rm mean}$, where $\rho_{\rm mean}$
is the mean matter density on large scales at the redshift of the
cluster.

\section{Cluster \cl}
\label{Sec:Cluster}

\cl, first reported in S12, was discovered in the IDCS as a striking
three-dimensional overdensity of massive galaxies with photometric
redshifts at $\zphot \approx 1.8$ (Figure \ref{Fig:sza}, upper right
panel).  Follow-up spectroscopy, with \hst/WFC3 in the infrared and
Keck/LRIS in the optical, identified seven robust spectroscopic
members at $z=1.75$ and ten additional likely members with less
certain redshifts.  The \hst\ data also revealed the presence of a
strong gravitational arc, the analysis and implications of which are
discussed in a companion paper \citep{gonzalez12}.

\cl\ is also detected in shallow \chandra\ data from \citet{murray05},
with an X-ray luminosity of $L_{\mbox{\scriptsize 0.5-2keV}} = (5.5
\pm 1.2) \times 10^{44}$ erg/s.  Using the $\Mfive$--L$_X$ scaling
relation of \citet{vikhlinin09} and the \citet{duffy08}
mass--concentration relation, this corresponds to a mass of $\Mtwoc
\simeq 5.6 \times 10^{14}$ \msun\ (S12)\footnote{In S12 the mass is
  measured with respect to a critical overdensity, rather than mean
  overdensity we use in this paper.}.
The \hst/ACS and WFC3 color image, along with the X-ray detection
(yellow contours) are shown in the lower panel of Figure
\ref{Fig:sza}.

While \cl\ appears from the X-ray data to be impressively massive,
particularly at such an extreme redshift, the X-ray luminosity is
based on only 54 cluster X-ray photons.  Further, it relies on a
high-scatter mass proxy, the $\Mfivec$-$L_X$ relation, which is only
calibrated at low ($z<1$) redshift.  The uncertainty of this X-ray
mass estimate is at least $\sim 40\%$.

\section{Data}
\label{Sec:Data}
\subsection{Sunyaev Zel'dovich Observations}
\cl\ was observed with the SZA, an 8-element radio interferometer
optimized for measurements of the SZ effect.  During the time in which
the data were obtained, the array was sited at the Cedar Flat location
described in \citet{culverhouse10}, and was in the ``SL''
configuration.  In this configuration, six elements comprise a compact
array sensitive to arcminute-scale SZ signals, and two outrigger
elements provide simultaneous discrimination for compact radio source
emission.  The compact array and outrigger baselines provide baseline
lengths of 0.35-1.3 k$\lambda$ and 2-7.5 k$\lambda$, respectively.

\begin{figure*}[bthp]
\epsscale{1.15} 
\plotone{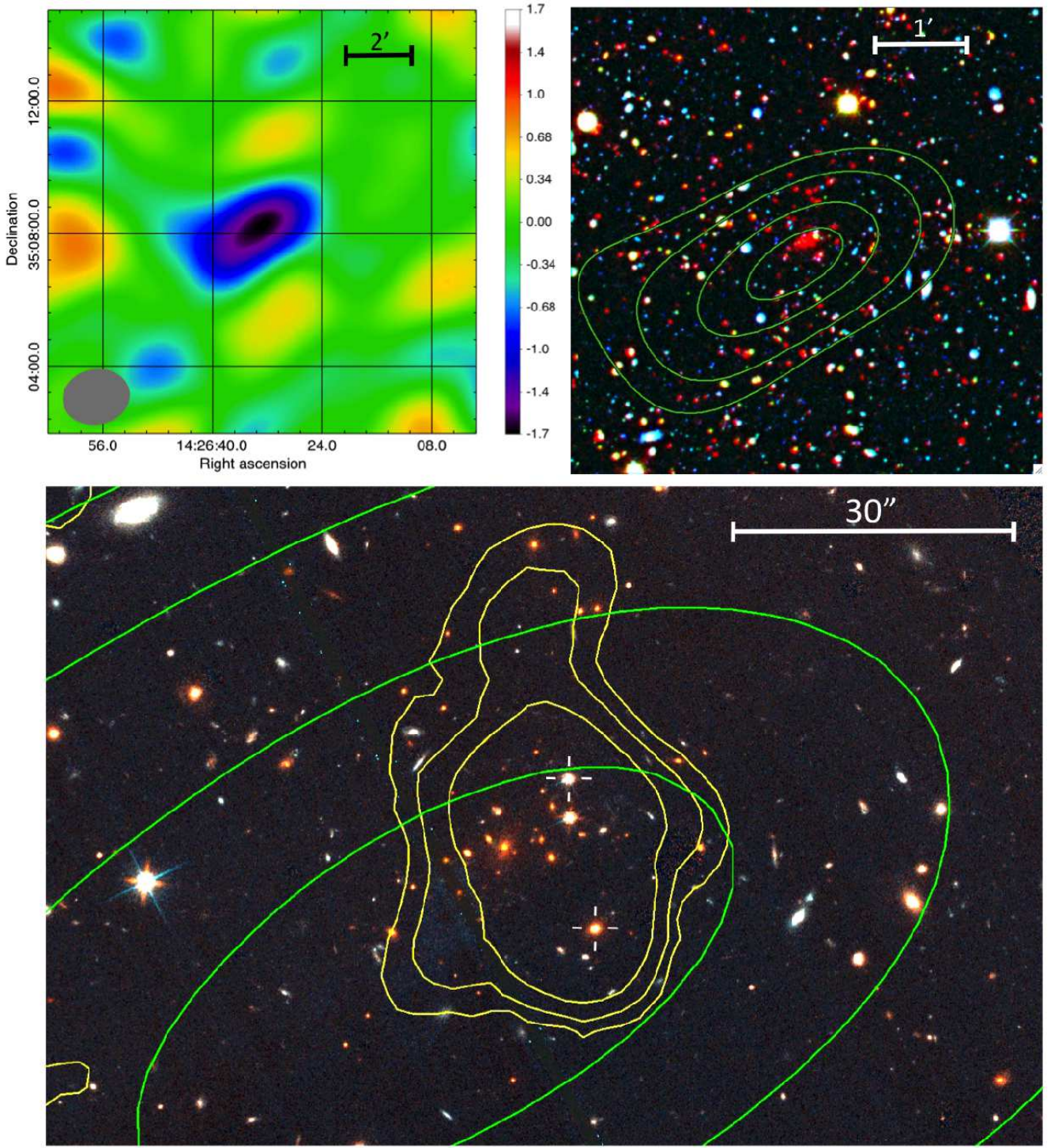}
\caption{\footnotesize {\it Upper left:} $12.8\arcmin \times
  12.8\arcmin$ SZA mm-wave map showing the SZ decrement of \cl.  The
  flux scale in mJy/beam is shown to the right of the map, and the
  FWHM of the CLEAN beam is shown in the lower left corner. {\it Upper
    right:} A $5\arcmin\ \times 5\arcmin$ $B_WR[4.5]$ pseudo-color
  image of \cl, with the SZ contours overplotted.  The optical and
  infrared data are from the NOAO Deep, Wide--Field Survey
  \citep[NDWFS;][]{ndwfs99} and SDWFS, respectively.  {\it Lower
    panel:} \hst\ F814W+F160W pseudo-color zoom--in of \cl.  The SZ
  contours (green) from the present work and X-ray contours (yellow)
  from S12, corresponding to 0.131, 0.077 and 0.044 counts per
  $2\arcsec \times 2\arcsec$ pixel in 8.3 ks in the 0.5-2~KeV band,
  are overplotted.  Two sources are indicated by hash marks.  The
  northern source is an X-ray luminous QSO that is also a cluster
  member; the southern source is a radio-bright AGN, the emission from
  which was removed from the SZA map.  Both are discussed further in
  the text.  In all panels North is up, East is left, and a scale bar
  is given.}
  \label{Fig:sza}
\end{figure*}

The cluster was observed for four sidereal passes on nights ranging
over UT 2011 June 9--29, for a total of 6.24~hours of unflagged,
on-source integration time.  Data were obtained in an 8~GHz passband
centered at 31~GHz.  The resulting noise level achieved by the short
(SZ-sensitive) baselines was 0.29~mJy/beam, or 22 $\mu$K$_{\rm RJ}$
with a synthesized beam of $118^{\prime\prime} \times
143^{\prime\prime}$.  For the long baselines, the noise level achieved
was 0.29~mJy/beam in a $19\farcs9 \times 14\farcs4$ beam.  The data
were reduced using the Miriad software package \citep{sault95}, and
the absolute calibration was determined using periodic measurements of
Mars calibrated against the \citet{rudy87} model.  A systematic
uncertainty of 5\% has been applied to the SZ measurements due to the
uncertainty in the absolute calibration.  The SZA map is shown in the
upper left panel of Figure \ref{Fig:sza}.

\subsection{Comptonization and Mass of \cl}
\label{sec:sza_analysis}

To determine the properties of \cl\ from the SZA data, we make use
of the fact that the integrated Compton $y$-parameter scales with
total mass.  Assuming a spherically symmetric pressure distribution
within the cluster, the integrated $y$-parameter is given by
\begin{equation}
\label{eqn:yint}
YD_A^2=\frac{\sigma_T}{m_ec^2}\int P(r) dV,
\end{equation}
where $D_A$ is the angular diameter distance, $\sigma_T$ is the
Thomson cross-section of an electron, $m_e$ is the electron mass, $c$
is the speed of light, $P(r)$ is the pressure profile, and the
integral is over a spherical volume centered on the cluster
\citep[see, e.g.,][]{marrone11}.  The pressure profile $P(r)$ is
determined by fitting a model profile to the SZA data.  We adopt the
generalized \citet{nfw} model proposed by \citet{nagai07}:
\begin{equation}
\label{eqn:gnfw}
P(r) = \frac{P_0}{x^\gamma (1+x^\alpha)^{(\beta-\gamma)/\alpha}},
\end{equation}
where $x=r/r_s$.  We fix the power law exponents $(\alpha, \beta,
\gamma)$ to the ``universal'' values reported by
\citet{arnaud10}, and leave $P_0$ and $r_s$ free to
vary\footnote{\citet{arnaud10} suggest other power law exponents
  appropriate for clusters with a known morphology, but
  \citet{marrone11} find that these alternative choices make a
  $\lesssim 5$\% difference in the estimate of $Y_{500}$.}.

One important source of systematic errors in SZ cluster measurements
is radio point sources.  These synchrotron sources are often variable,
are known to be correlated with clusters, and can fill in the SZ
decrement at low frequencies \citep[see, \eg][]{carlstrom02}.  Wide
angular scale surveys such as the SPT survey generally mask regions
where bright point sources are detected, which changes the survey
selection function but can be accounted for using simulations
\citep{vanderlinde10}.  However, for targeted SZ follow-up of sources
selected in other bands, point source fluxes must be measured and
removed in order to accurately estimate the cluster mass.  Due to the
temporal variability of these sources, the point source fluxes ideally
should be determined concurrently with the SZ measurement.  The SZA
uses its longer outrigger baselines to perform such simultaneous point
source measurements.  For spatially compact sources, the emission
measured on the outrigger baselines (smaller angular scales) can be
assumed to be representative of the emission measured on the short
SZ-sensitive baselines (larger angular scales), and can thereby be
removed from the SZ signal \citep[see, \eg][]{muchovej07}.

Two such point sources were discussed in S12 and are indicated by hash
marks in Figure \ref{Fig:sza}.  The northern source is an X-ray bright
QSO that is also a spectroscopically confirmed member of \cl.  As
described in S12, the contribution to the cluster's X-ray flux from
this QSO had to be removed to obtain a reliable cluster mass. This QSO
is not detected in the radio, however, with a formal 31 GHz flux
density of $-0.04 \pm 0.31$ mJy, and therefore it has no impact on the
measured cluster mass.  A second point source, bright enough in the
radio to fill in the SZ decrement of \cl, was identified near the
cluster center at $14^{\rmn{h}}26^{\rmn{m}}32\fs1$, $35^\circ
08\arcmin 15\farcs2$, indicated by the southern hash mark in Figure
\ref{Fig:sza}.  This source corresponds to a $\sim 95$ mJy source
identified at 1.4 GHz in the NVSS \citep{condon98} and FIRST
\citep{becker95} catalogs.  The lower--frequency radio survey data
indicate that this source is unresolved at the angular scales to which
the SZA is sensitive, allowing it to be modeled as a point source and
constrained using the long baseline data.  The flux density of this
source is left as a free parameter in the cluster model fit, and is
found to be $5.3 \pm 0.3$~mJy.  The resulting spectral index of $-0.93
\pm 0.02$ is consistent with synchrotron radiation.

A Markov Chain Monte Carlo (MCMC) fit is used to jointly constrain
$P_0$, $r_s$, and the flux densities of the two point sources in our
fiducial cosmology.  The results of this fit are used to compute the
integrated $y$-parameter and the total cluster mass.  By convention,
both the total cluster mass and the integrated $y$-parameter are
reported within a radius corresponding to a fixed overdensity
$\Delta=500$ relative to the critical density of the Universe at the
redshift of the cluster. The integrated quantities are determined by
finding the overdensity radius $r_{500}$ (equivalent to the mass,
since $M_\Delta = (4\pi r_\Delta^3 /3) \Delta \rho_c(z)$) and
integrated $y$-parameter $Y_{500}$ that are mutually consistent with
the \citet{andersson11} $M_{500}$--$Y_{500}$ scaling relation.  We
choose this relation since the clusters with which it is measured have
the highest mean redshift.  The self-consistent values of $r_{500}$
and $Y_{500}$ are determined iteratively at each step in the MCMC, and
the scaling relation parameters are sampled at each step using their
reported means and variances.  Using this method, we find that
$\Mfivec = (2.6 \pm 0.7) \times 10^{14}$ \msun, where the error is
statistical in nature.  Assuming the \citet{duffy08} concentration
relation, this corresponds to $\Mtwoc = (4.1 \pm 1.1) \times 10^{14}$
\msun. This mass agrees well with the X-ray mass of $\Mtwoc \simeq 5.6
\times 10^{14}$ \msun, for which the error is at least 40\%.

A significant uncertainty, inherent to both SZ and X-ray measurements
of the masses of high redshift clusters, is due to the fact that no
published mass-observable scaling relation has been estimated using a
sample containing clusters at redshifts significantly greater than
unity.  While \citet{andersson11} found no compelling evidence for
redshift evolution in the $M_{500}$--$Y_{500}$ scaling relation out to
$z \sim 1$, their sample consisted of lower-redshift and higher-mass
objects whose scaling properties may be systematically different from
those of \cl.

To enable the SZ mass of \cl\ to be easily calculated with future
scaling relations, we also report the spherically-averaged
dimensionless Comptonization, $Y_{\mbox \scriptsize \rm sph,500} =
(7.9 \pm 3.2) \times 10^{-12}$.  This corresponds to a gas mass at
$\rfive$ of $M_{\mbox \scriptsize \rm gas} = (2.9 \pm 1.2) \times 10^{13}$
\msun\ for a constant temperature of 5 keV.  This temperature is
consistent with the X-ray measurement (S12); for a cooler 4 keV
cluster the inferred gas mass would increase by 25\%. 

To assess the SZ detection significance, we fit a cluster model to the
visibility data and compare to a null model.  Our model consists of an
elliptical Gaussian SZ decrement located at the X-ray centroid along
with the nearby radio point source, and our null model consists only
of the point source.  We find a $\Delta \chi^2$ of $37.65$ between the
model and null model fits.  Since the (fixed-position) elliptical
Gaussian adds four degrees of freedom, this $\Delta \chi^2$
corresponds to a detection significance of 5.3 $\sigma$, or a false
detection probability of $1.32 \times 10^{-7}$.  Because the SZ signal
was sought only at the position of the X-ray emission, this detection
is very secure.  In contrast, a false 5.3 sigma detection is far more
plausible in a wide--field survey like SPT because there are of order
$10^6$ independent pixels in which such a deviation can arise by
chance.
We therefore consider the pointed, follow-up detection of \cl\
to be robust.
 
If the cluster position is not fixed to the X-ray centroid, the
best--fit centroid of the SZ signal is
$14^{\rmn{h}}26^{\rmn{m}}34\fs0$, $35^\circ 08\arcmin 02\farcs6$,
which is slightly offset $(\approx 32\arcsec$ and $25\arcsec)$ from
the X-ray and BCG positions, respectively.  As the uncertainty in the
SZ centroid is $\approx 35\arcsec$, this offset is not statistically
significant.  Absent the X-ray positional prior, the significance of
the SZ detection is almost identical (5.2 $\sigma$).

\section{Probability of Existence in a $\Lambda$CDM Universe}
\label{Sec:Rarity}

While not as massive as the $z\sim 1$ SPT clusters, \cl, by virtue of
its exceptional redshift and substantial mass, is arguably one of the
more extreme systems discovered to date.  It is therefore interesting
to calculate the probability of discovering it in the IDCS in a
standard $\Lambda$CDM Universe.

We use the exclusion curve formalism presented in \citet[][hereafter
M11]{mortonson11}, which tests whether a single cluster is rare enough
to falsify $\Lambda$CDM and quintessence at a given significance
level, accounting for both sample and parameter variance.  The M11
formalism uses cluster masses measured within a sphere defined by an
overdensity 200 times the mean, rather than critical, density of the
Universe.  In this convention, adopting the \citet{duffy08}
mass--concentration relation, \cl\ has a mass of $\Mtwom = (4.3 \pm
1.1) \times 10^{14}$ \msun.

The M11 formalism requires that observed cluster masses be corrected
for Eddington bias caused by the steep slope of the cluster mass
function at this mass scale and redshift.  We follow the prescription
given in Appendix A of M11:
\begin{equation}
\ln M=\ln M_{\mbox{\scriptsize obs}}+\frac{1}{2}\gamma\sigma^2_{\ln M},
\end{equation}
where $M$ is the expected value for the true mass given the observed
mass, $\gamma$ is the local slope of the mass function, and $\sigma$
is the uncertainty associated with the observations. Here we use the
fitting function for $\gamma$ provided in M11, which yields
$\gamma=-5.8$. We thus derive that the expected value for this
bias-corrected mass is $\Mtwom = (3.6 \pm 0.9) \times 10^{14}$ \msun.

\begin{figure}[bthp]
\epsscale{1.15} 
\plotone{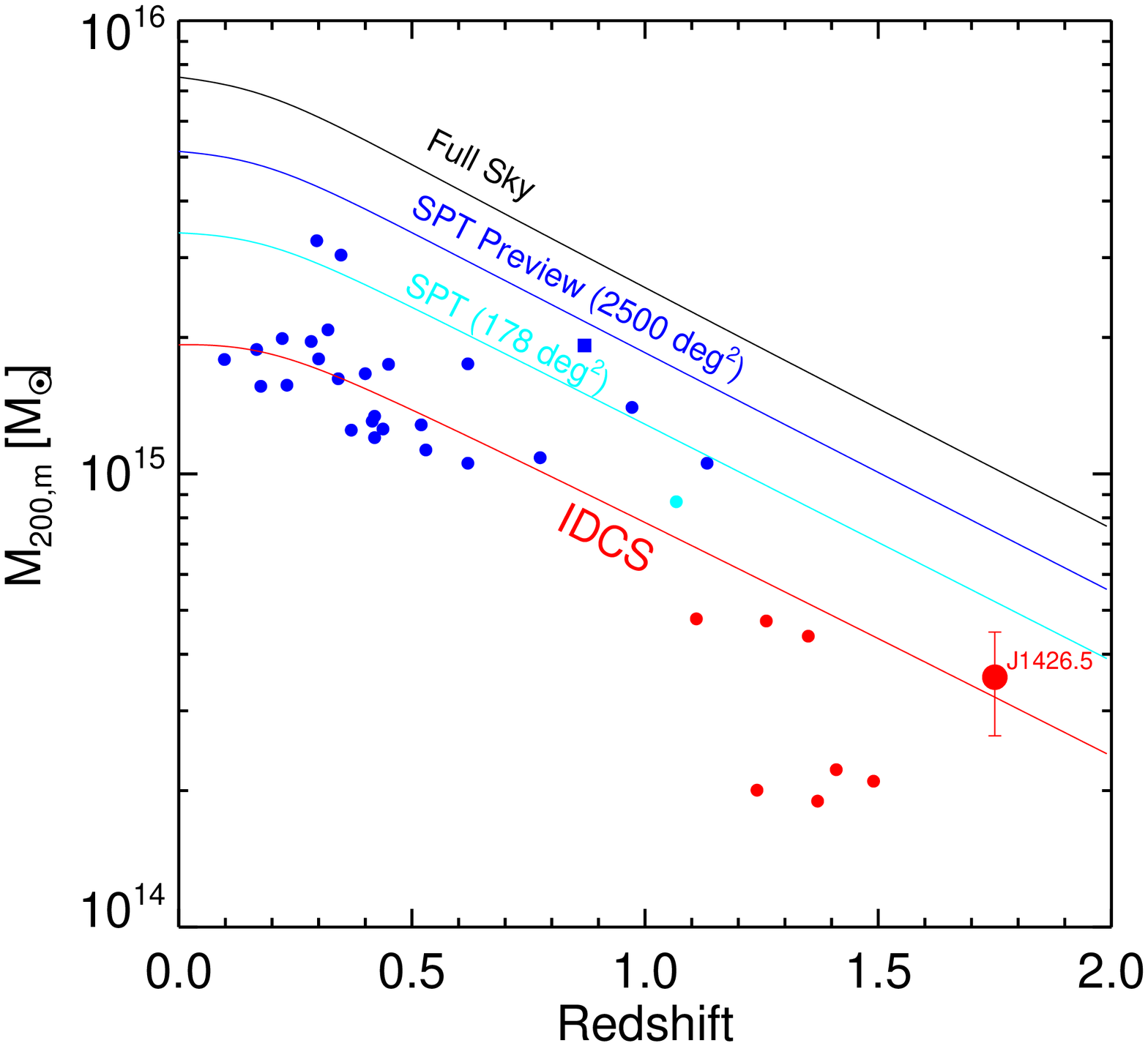}
\caption{M11-style plot showing the mass and redshift of \cl\ (large
  red circle), along with other $z>1$ ISCS clusters
  (\citealt{brodwin11,jee11}, small red circles) and SPT clusters
  (\citealt{williamson11}, small blue circles; \citealt{brodwin10},
  small cyan circle).  The solid red curve is the 95\% exclusion curve
  for the IDCS area.  The cyan, blue and black curves are the
  exclusion curves for the currently published full depth SPT survey
  (178 deg$^2$, \citealt{vanderlinde10}), the 2500 deg$^2$ SPT preview
  survey of \citet{williamson11} and the full sky, respectively.  The
  square symbol represents cluster SPT-CL J0102-4915, first reported
  as ACT-CL J0102-4915 \citep{menanteau10}. We plot this cluster at
  the spectroscopic redshift of $z=0.870$ reported in
  \citet{menanteau12}.  All clusters are color-coded to the
  appropriate exclusion curve.}
 \label{Fig:rarity}
\end{figure}

In Figure \ref{Fig:rarity} we plot the appropriately bias-corrected
mass for \cl\ along with 95\% confidence exclusion curves using the
M11 prescription. For comparison we include rare, massive clusters
from the SPT survey \citep{williamson11,brodwin10} and other massive
$z>1$ clusters from the ISCS for which weak lensing or X-ray masses
are available \citep{brodwin11,jee11}.  All masses are deboosted in
accordance with the M11 prescription.

The lowest curve in the figure corresponds to the 95\% exclusion curve
for clusters within the 8.82 deg$^2$ IDCS area. The central value of
the deboosted SZ mass is above the curve, formally falsifying
$\Lambda$CDM at the 95\% level.  We note, however, that the
statistical error bar overlaps the exclusion curve.  Furthermore,
uncertainties in the scaling relation could easily lower the mass
below the exclusion curve.

Rather than asking whether such a cluster should have been detected in
the IDCS, perhaps a fairer question is to ask whether such a cluster
should have been detected in the full area of sky covered by all
cluster surveys capable of detecting such a cluster.  Numerous X-ray
surveys of varying areas and depths could have detected \cl, but for
simplicity we consider the SPT survey, which has the sensitivity to
detect clusters like \cl, even at $z=1.75$ \citep{vanderlinde10}.  The
first 178 deg$^2$ of the SPT survey does not contain a single cluster
comparable to \cl\ in mass and redshift \citep{vanderlinde10}, which
suggests that the true abundance of such clusters is far lower than
our discovery would indicate.  Neither the Atacama Cosmology Telescope
(ACT) project nor the {\it Planck} project surveys are sensitive
enough to detect \cl\ \citep{marriage11,planck11}.  If we consider
\cl\ to have been drawn from an area of at least $\sim178$ deg$^2$
(shown as the cyan curve in Fig.~\ref{Fig:rarity}) rather than the
IDCS volume, then the cluster is no longer in contradiction with
$\Lambda$CDM; we were simply somewhat lucky to detect it in the
comparatively tiny ($8.82$ deg$^2$) IDCS area.

\section{Evolution to the Present Day}
\label{Sec:Evolution}

Although the mere existence of \cl\ may not have significant
cosmological implications, it is nevertheless among the rarest, most
extreme clusters ever discovered.  As such, it is interesting to
understand its nature in the context of the growth of the largest
structures.

\begin{figure*}[bthp]
\epsscale{1.15} 
\plotone{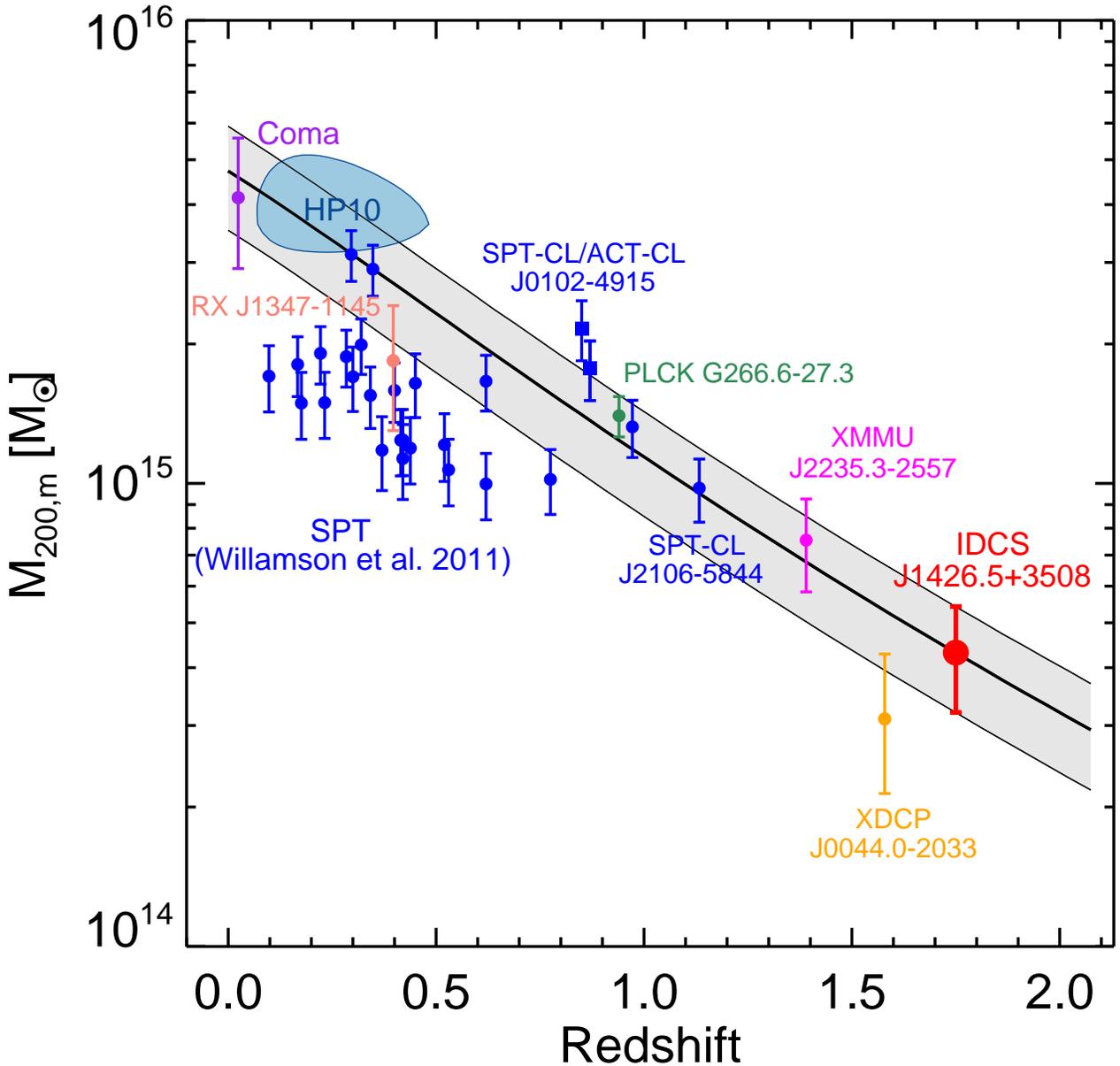}
\caption{Predicted mass growth of \cl\ vs.\ redshift based on
  abundance matching.  \cl\ is the large red circle and the predicted
  growth in its mass is shown as a thick black line.  The 1 $\sigma$
  errors, stemming from the mass measurement errors, are shown as the
  shaded region.  An assortment of the rarest, most massive clusters
  found at all redshift is shown for comparison and discussed in the
  text.  SPT-CL J0102-4915 was first reported as ACT-CL J0102-4915 by
  \citet{menanteau10} and \citet{marriage11}.  The independent mass
  measurements from both surveys are plotted as the square symbols for
  this cluster at the spectroscopic redshift ($z=0.8701$) reported in
  \citet{menanteau12}.  \cl\ is consistent with being a member of the
  most extreme population of virialized structures in the Universe.}
 \label{Fig:evolution}
\end{figure*}

Using the \citet{tinker08} mass function, we first identify the space
density of clusters at $z=1.75$ that have masses greater to or equal
to that of \cl.  At each redshift we then associate the clusters with
that space density with its descendents.  In Figure
\ref{Fig:evolution} we plot the mass evolution of \cl\ (large red
circle) from $z=1.75$ to the present day (thick black line).  For
comparison we also plot a representative selection of the most extreme
clusters found to date at all redshifts.  This abundance--matching
calculation is designed to predict the mean expected growth of \cl.
We note that there are intrinsic variations in accretion histories due
to the stochastic nature of mergers \citep[\eg][]{wechsler02}, but
consider a full treatment beyond the scope of the present work.

As demonstrated in this figure, \cl\ is much more than just another
massive, very high redshift cluster.  It is an evolutionary precursor
to the most massive known clusters at all redshifts.  Indeed, its
descendent population consists of the most massive virialized objects
ever discovered.

There are no other confirmed clusters at $z>1.75$ with masses greater
than $10^{14}$ \msun.  Between $1.5 < z < 1.75$ only one cluster, XDCP
J0044.0-2033 at $z=1.58$ \citep{santos11, fassbender11}, is comparable
to \cl, though it is somewhat less massive and below the growth trend.

The most massive $z>1$ X-ray selected cluster, XMMU J2235.3-2557 at
$z=1.39$ \citep{mullis05,rosati09}, falls along the evolutionary path
of \cl, as does the most massive $z>1$ cluster in the SPT survey,
SPT-CL 2106-5844 at $z=1.13$ \citep{foley11, williamson11}. Similarly,
the only $z\sim 1$ {\it Planck} cluster reported to date, PLCK
G266.6-27.3 at $z=0.94$ is also consistent.

SPT-CL J0102-4915, first reported as ACT-CL J0102-4915 from the ACT
survey \citep{menanteau10,marriage11}, is among the most significant
detections overall in both of these SZ surveys.  Masses of this
cluster from the two SZ surveys are plotted as the square symbols at
the spectroscopic redshift of $z=0.870$ \citep{menanteau12}. The lower
SPT mass is measured from the SZ decrement, whereas the higher ACT
mass (slightly offset in redshift for clarity) is combined from
several methods.  The masses are consistent within the errors.  \cl\
is consistent with growing into a cluster like this as well.  RX
J1347-1145 \citep{allen02,lu10} is the most luminous X-ray cluster in
the sky, though it is not quite massive enough to be considered a
descendent of \cl.  In the local Universe the Coma cluster represents
the kind of extreme cluster \cl\ will evolve into.

\citet{holz10} recently predicted the properties of the single most
massive cluster that can form in a $\Lambda$CDM Universe with Gaussian
initial conditions and where the growth of structure follows the
predictions of general relativity.  They predict this cluster should
be found at $z=0.22$ and have a mass of $\Mtwom = 3.8 \times 10^{15}$
\msun.  Their 68\% contour in predicted mass and redshift is shown as
a light blue contour and labeled HP10 in the figure.  \cl\ is
completely consistent with being an evolutionary precursor to the most
massive halo predicted to exist in the observable Universe.

\section{Discussion}
\label{Sec:Discussion}
 
Given its extreme mass it is not surprising that \cl\ was detected in
shallow (5 ks) \chandra\ images (S12), despite its high redshift.  The
robust, high-significance SZA detection reported herein confirms that
\cl\ is one of the most massive collapsed structures evolving in the
Universe.  Although it was discovered in a small ($\approx 9$ deg$^2$)
infrared-selected cluster survey, the odds of finding such a rare,
massive cluster were quite low, $0.8\%$ ($0.2\%$) using the
\citet{holz10} formalism with the deboosted (measured) mass.  Finding
the precursors of \cl\ at even earlier times will require much larger
surveys.

In this section we examine why the current generation of SZ surveys,
despite covering areas ranging from 10$^2$--10$^4$ deg$^2$, have not
yet seen a cluster as extreme as \cl\ at $z>1.5$.  We also discuss the
potential of the next generation of SZ surveys to discover its
evolutionary precursors.

\subsection{Current SZ surveys}

The detectability of a cluster like \cl\ in an SZ cluster survey
depends on three factors.  The survey must have sufficient mass
sensitivity at high redshift, it must cover sufficient area so that
the expectation value of the number of clusters is greater than unity,
and the clusters in question must not contain bright radio point
sources nor have any along the line of sight.

In terms of sensitivity, only the SPT survey is currently capable of
detecting \cl\ at this high redshift. \citet{vanderlinde10} report a
formal 50\% completeness of $\Mtwom \approx 3.5 \times 10^{14}$ \msun\
at $z=1.5$, and this improves to $\ga 90\%$ at $z=1.75$.  That mass
limit is below both the measured and deboosted mass of \cl.  So the
SPT should nominally have detected such a cluster were it present at
$z=1.75$ in the single-band analysis of their first 178 deg$^2$
\citep{vanderlinde10}.  However, since the SPT completeness limit at
high redshift is uncertain at the 30\% level in mass
\citep{vanderlinde10}, the non-detection of a single high--redshift
cluster like \cl\ is unremarkable.

A more interesting question, related to the second criterion and to
the rarity analysis of \textsection{\ref{Sec:Rarity}}, is that of how
many such clusters are actually expected in the current SPT area?
According to the \citet{holz10} formulation, we expect only 0.2 such
clusters in 178 deg$^2$, and therefore none should have been found.
In the complete, full-depth 2500 deg$^2$ SPT survey, \citet{holz10}
predict the presence of 2.4 such clusters, though shot noise will play
a role in whether any are found.

Finally, as discussed in \textsection{\ref{sec:sza_analysis}}, \cl\
has a radio--bright AGN at $z=1.535$ along the line-of-sight,
indicated by the southern hash marks in Figure \ref{Fig:sza}.  At a
redshift separation of $\Delta z>0.2$, it is not physically associated
with the cluster. Its radio emission was constrained using the long
SZA baselines, which are insensitive to the arcminute angular scale
SZ decrement.  If the measured 1.4 to 31 GHz spectral index of this
source ($-0.93$) is valid to 150~GHz, the AGN flux would be 1.2 mJy at
this frequency, well below the 5 mJy threshold for masking individual
point sources in the SPT survey.  However, it would partially fill in
the $\sim 14$ mJy SZ decrement at 150 GHz, reducing the SPT detection
significance and biasing the mass estimate low by $\sim 10 - 15\%$.

It is possible that the AGN content of clusters at the highest
redshifts may be substantially larger than at intermediate ($z<1$)
redshifts.  Indeed, \citet{galametz09} find evidence for a strong
increase in the incidence of AGN in clusters with redshift from $0.2 <
z < 1.4$, particularly for X-ray and IR-selected AGN.  The evidence
for rapid number density evolution in radio-selected AGN is not as
conclusive.  In some surveys \citep[\eg][]{galametz09} it increases as
quickly in clusters as in the field, whereas others
\citep[\eg][]{gralla11} find it is roughly constant.  One major
advantage of CARMA for targeted SZ observations of clusters is the
ability to simultaneously identify and remove contamination from such
(generally variable) sources, as in this work, and thus recover
accurate cluster masses.

\subsection{Next--Generation SZ surveys}

Several next--generation SZ experiments designed for CMB polarization
measurements, such as SPTpol \citep{SPTpol} and ACTpol \citep{ACTpol},
are currently being deployed.  These experiments will possess typical
SZ mass sensitivities a factor of several lower than present surveys.
Here we calculate the expected yield of high--redshift clusters
similar to \cl\ for a canonical 500 deg$^2$ next--generation survey
with constant $\Mtwom$ completeness limits of $1.2 \times 10^{14}$
\msun\ and $1.5 \times 10^{14}$ \msun\ at the 50\% and 90\% levels,
respectively.

\begin{figure}[bthp]
\epsscale{1.15} 
\plotone{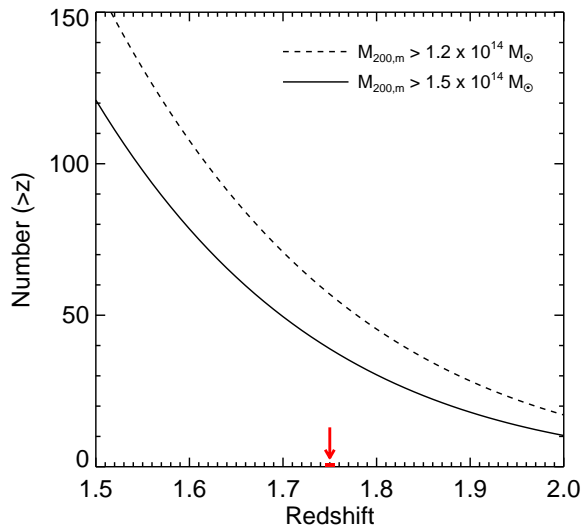}
\caption{Predicted cumulative counts of massive $1.5 < z < 2$ clusters
  at the 50\% ($\Mtwom = 1.20 \times 10^{14}$ \msun) and 90\% ($\Mtwom
  = 1.5 \times 10^{14}$ \msun) completeness limits of a representative
  500 deg$^2$ next--generation SZ survey.  The curves account for
  losses due to incompleteness.  The arrow indicates the redshift of
  \cl.}
  \label{Fig:nextgen}
\end{figure}

The same methodology that explains the lack of a cluster like \cl\ in
the current SPT survey area predicts that a significant number of very
high redshift clusters, somewhat less massive than \cl, will be found
with the polarization experiments.  Figure \ref{Fig:nextgen} shows the
expected cumulative yield, based on the \citet{holz10} formulation, at
the two mass limits given above, accounting for the expected
completenesses.   

The combination of area and depth of the next generation SZ surveys
will allow them to identify a large statistical sample of massive, hot
clusters at $1.5 < z < 2.0$.  As such they provide a natural extension
of the infrared surveys, which have characterized galaxy clusters to
$z=1.5$.  A 100 deg$^2$ region that will be covered by both SPTpol and
ACTpol will also contain \spitzer\ infrared data, from an ongoing
\spitzer\ Exploration Class program (PID 80096, PI Stanford).  This
will allow optimal high-redshift cluster detection and study at both
IR and mm wavelengths, the only two cluster probes that are nearly
redshift-independent at $1<z<2$.

Despite the wealth of new, high redshift clusters to be found by the
next-generation SZ surveys, there are only even odds ($p\sim0.5$) of
finding a $z=2$ precursor to \cl.  Unless, of course, we are fortunate
once again.

\section{Conclusions}
\label{Sec:Conclusions}

We have presented a robust 5.3 $\sigma$ SZ detection of the
infrared-selected cluster \cl\ at $z=1.75$, by far the most distant
cluster with an SZ detection to date.  Using the \citet{andersson11}
$M_{500}$--$Y_{500}$ scaling relation, we find $\Mtwom = (4.3 \pm 1.1)
\times 10^{14}$ \msun, in agreement with the X-ray mass reported in
S12.

The chance of finding \cl\ in the $\sim$ 9 deg$^2$ IDCS was $\la 1\%$,
which, taken at face value, marginally rules out $\Lambda$CDM.  Within
the much larger total area probed by other SZ and X-ray surveys, its
existence is not unexpected.

In a $\Lambda$CDM cosmology, with the growth of structure as predicted
by general relativity, \cl\ will grow to a mass of $\Mtwom \approx 4.7
\times 10^{15}$ \msun\ by the present day.  It therefore shares an
evolutionary path with the most massive known clusters at every
redshift.  Along with several of the rarest, high-redshift clusters
currently known, \cl\ is consistent with evolving into the single most
massive cluster expected to exist in the observable Universe.

\acknowledgments Support for CARMA construction was derived from the
states of California, Illinois, and Maryland, the James S. McDonnell
Foundation, the Gordon and Betty Moore Foundation, the Kenneth T. and
Eileen L. Norris Foundation, the University of Chicago, the Associates
of the California Institute of Technology, and the National Science
Foundation. Ongoing CARMA development and operations are supported by
the National Science Foundation under a cooperative agreement, and by
the CARMA partner universities.  The work at Chicago is supported by
NSF grant AST-0838187 and PHY-0114422.

This work is based in part on observations made with the {\it Spitzer
  Space Telescope}, which is operated by the Jet Propulsion
Laboratory, California Institute of Technology under a contract with
NASA. Support for this work was provided by NASA through an award
issued by JPL/Caltech.  This work is based in part on observations
obtained with the {\it Chandra X-ray Observatory} (CXO), under
contract SV4-74018, A31 with the Smithsonian Astrophysical Observatory
which operates the CXO for NASA.  Support for this research was
provided by NASA grant G09-0150A.  Support for \hst\ programs 11663
and 12203 were provided by NASA through a grant from the Space
Telescope Science Institute, which is operated by the Association of
Universities for Research in Astronomy, Inc., under NASA contract NAS
5-26555.  This work is based in part on data obtained at the
W.~M.~Keck Observatory, which is operated as a scientific partnership
among the California Institute of Technology, the University of
California and the National Aeronautics and Space Administration.  The
Observatory was made possible by the generous financial support of the
W.~M.~Keck Foundation.  This work makes use of image data from the
NOAO Deep Wide--Field Survey (NDWFS) as distributed by the NOAO
Science Archive. NOAO is operated by the Association of Universities
for Research in Astronomy (AURA), Inc., under a cooperative agreement
with the National Science Foundation.

We thank the anonymous referee for suggestions which improved the
manuscript, Bradford Benson for helpful discussions, F.~Will High and
Keith Vanderlinde for providing published SPT masses in a digital
form, Matt Ashby for creating the IRAC catalogs for SDWFS, Michael
Brown for combining the NDWFS with SDWFS catalogs, Alexey Vikhlinin
for advice on the analysis of the \chandra\ data, and Daniel Holz for
providing his predictions in an electronic format.  This paper would
not have been possible without the efforts of the support staffs of
CARMA, the Keck Observatory, {\it Spitzer Space Telescope}, {\it
  Hubble Space Telescope}, and {\it Chandra X-ray Observatory}.
Support for MB was provided by the W.~M.~Keck Foundation.  AHG
acknowledges support from the National Science Foundation through
grant AST-0708490.  The work by SAS at LLNL was performed under the
auspices of the U.~S.~Department of Energy under Contract
No. W-7405-ENG-48.

\bibliographystyle{astron2}
\bibliography{bibfile}
\end{document}